\documentclass[journal=jacsat,manuscript=article]{achemso}
\setkeys{acs}{maxauthors=3, etalmode=truncate}
\usepackage{xr}
\makeatletter
\newcommand*{\addFileDependency}[1]{
  \typeout{(#1)}
  \@addtofilelist{#1}
  \IfFileExists{#1}{}{\typeout{No file #1.}}
}
\makeatother

\newcommand*{\myexternaldocument}[1]{
    \externaldocument{#1}
    \addFileDependency{#1.tex}
    \addFileDependency{#1.aux}
}

\myexternaldocument{Supportinginformation}

\usepackage[version=3]{mhchem} 
\usepackage[T1]{fontenc} 
\usepackage{gensymb}
\usepackage{xcolor}

\author{Han Gao}
\affiliation[BAQIS]
{Beijing Academy of Quantum Information Sciences, Beijing 100193, China}
\alsoaffiliation[EqualContribution]
{These authors contribute equally}

\author{Jian-Huan Wang}
\affiliation[BAQIS]
{Beijing Academy of Quantum Information Sciences, Beijing 100193, China}
\alsoaffiliation[EqualContribution]
{These authors contribute equally}

\author{Ji-Yin Wang}
\email{wang_jy@baqis.ac.cn}
\affiliation[BAQIS]
{Beijing Academy of Quantum Information Sciences, Beijing 100193, China}

\author{Jian-Jun Zhang}
\email{jjzhang@iphy.ac.cn}
\affiliation[IOP]
{Beijing National Laboratory for Condensed Matter Physics and Institute of Physics,
Chinese Academy of Sciences, Beijing 100190, China.
}

\author{Hongqi Xu}
\email{hqxu@pku.edu.cn}
\affiliation[BAQIS]
{Beijing Academy of Quantum Information Sciences, Beijing 100193, China}
\alsoaffiliation[PKU]
{Beijing Key Laboratory of Quantum Devices and School of Electronics, Peking University, Beijing 100871, China}

\title{Supercurrent and multiple Andreev reflections in Ge hut nanowire Josephson Junctions}

\begin{document}
\begin{center}
(Dated: \date{\today})
\end{center}
\newpage







\begin{abstract}
We report an experimental study of induced superconductivity in Ge hut nanowire Josephson junctions. The Ge hut nanowires are grown on prepatterned SiGe ridges via molecular beam epitaxy (MBE) and Josephson junction devices are fabricated by contacting the nanowires with Al electrodes. Low-temperature current-bias transport measurements of the Josephson junctions are performed and the measurements show that the devices exhibit gate-tunable supercurrent and excess current. The analysis of excess current indicates that the transparency of the Ge hut nanowire Josephson junctions is as high as 85\%. Voltage-bias spectroscopy measurements of the devices show multiple Andreev reflections up to the fourth order. With magnetic field and temperature-dependent measurements of the multiple Andreev reflections, the critical field and the critical temperature of the induced superconductivity in the Josephson junctions are extracted to be $\sim0.12\,\mathrm{T}$ and $\sim1.4\,\mathrm{K}$. The success in introducing superconductivity into Ge hut nanowires will stimulate their applications in building advanced quantum processors.
\end{abstract}

 Key words:\ Ge hut nanowire, Josephson junction, Supercurrent, Multiple Andreev reflection

\newpage

Recently emerged germanium (Ge) nanostructures obtained by epitaxial growth have become a primary platform for building state of the art quantum computation processors,\cite{Scappucci2021_NRM} including spin qubits,\cite{Hendrickx2020_Nature,Hendrickx2021_Nature,Jirovec2021_NM,Wang2024_Science,Froning2021_NN,Wang2022_NC} gate-controllable superconducting qubits\cite{Sagi2024_NC,Kiyooka2025_NanoLett,Zhuo2023_npj,Zheng2024_NanoLett} and qubits encoded in topological states.\cite{Maier2014_PRB,Adelsberger2023_PRB,Luethi2023_PRB,Laubscher2024_PRB} This fast development is fueled by the intrinsic properties of holes in Ge, such as strong spin-orbit interaction,\cite{Bulaev2007_PRL,Moriya2014_PRL} weak hyperfine interaction with nuclei, and high carrier mobility.\cite{Lodari2022_APL,Kong2023_AMI}  Various advanced quantum processors have been built with Ge nanostructures by taking the advantages of these properties. For instance, spin qubits with gate fidelity exceeding the error-correction thresholds have been built with the use of Ge two-dimensional heterostructures,\cite{Wang2024_Science} and ultra-fast spin qubit manipulations have been achieved in Ge core-shell nanowires\cite{Froning2021_NN} and Ge hut nanowires.\cite{Wang2022_NC,Liu2023_NanoLett} Superconducting quantum devices have also been developed by combining Ge nanostructures with superconductors\cite{Hendrickx2018_NC,Hendrickx2019_PRB,Vigneau2019_NL,Aggarwal2021_PRR,Tosato2023_CM,Valentini2024_NC,Lakic2025_NM,Ridderbos2020_NL,Wu2024_APL} for potential uses in building advanced quantum hardware. For example, gate-tunable superconducting qubits have been manufactured from Ge quantum wells\cite{Sagi2024_NC,Kiyooka2025_NanoLett} and Ge nanowires\cite{Zhuo2023_npj,Zheng2024_NanoLett}. Currently, Ge nanostructures proximitized by superconductors are considered to be an intriguing type of system to construct topologically protected Majorana zero modes.\cite{Maier2014_PRB,Adelsberger2023_PRB,Luethi2023_PRB,Laubscher2024_PRB} Overall, Ge nanostructures have become an indispensable material system in the field of modern quantum computing technology developments.

Among Ge nanostructures, Ge hut nanowires have garnered increasing interest. These Ge nanowires commonly have a low symmetric triangular cross section and thus possess a gate-tunable, ultra-strong hole spin-orbit interaction\cite{Liu2022_PRApplied,Liu2023_NanoLett} and allow sweet-spot qubit operations.\cite{Bosco2021_PRXQ,Bosco2021_PRL} Previously, the growth of Ge hut nanowires was achieved using molecular beam epitaxy (MBE)\cite{ZhangJJ2012_PRL,Gao2020_AM,Ming2023_Nanoscale} and the first Ge spin qubit was demonstrated on such nanowires.\cite{Watzinger2018_NC} However, in spite of these progresses, the high temperature employed during nanowire growth can cause Si-Ge interdiffusion.\cite{Baranov2006_PRB,Aqua2013_PhyRep} This could hinder the hole mobility in the nanowires and severely limit the potential of Ge hut nanowires as a versatile platform for building advanced quantum processors. Recently, we have realized the low temperature growth of Ge hut nanowires on prepatterned SiGe ridges using MBE\cite{Wang2025_NL} and demonstrated that these hut nanowires are purely Ge wires and exhibit a record high hole mobility. Thereby, constructing advanced quantum devices on these low-temperature MBE-grown Ge hut nanowires becomes feasible.

In this work, we present the first experimental realization of Josephson junctions on Ge hut nanowires grown by our low-temperature MBE technique. The devices are made by contacting the Ge hut nanowires with Al electrodes. Low-temperature transport measurements show gate-tunable supercurrent, and the transparency of Josephson junctions can be as high as 85\%. Multiple Andreev reflections (MARs) up to the fourth order are detected in voltage-biased spectroscopy measurements, from which the superconducting gap of the hybrids is extracted to be $\sim130\,\mathrm{\mu eV}$. With magnetic field- and temperature-dependent measurements, the critical magnetic field and critical temperature of the induced superconductivity in the devices are extracted to be $0.12\,\mathrm{T}$ and $1.4\,\mathrm{K}$, respectively. The results show that decent proximitized superconductivity has been achieved in Ge hut nanowires. Such an achievement is highly important for expanding the applications of Ge hut nanowires to different kinds of superconducting quantum devices, including, e.g., gatemon qubits and topological superconducting qubits, and cryogenic superconducting electronics such as superconducting diodes.

\begin{figure*}[!t]
\centering
\includegraphics[width=1\linewidth]{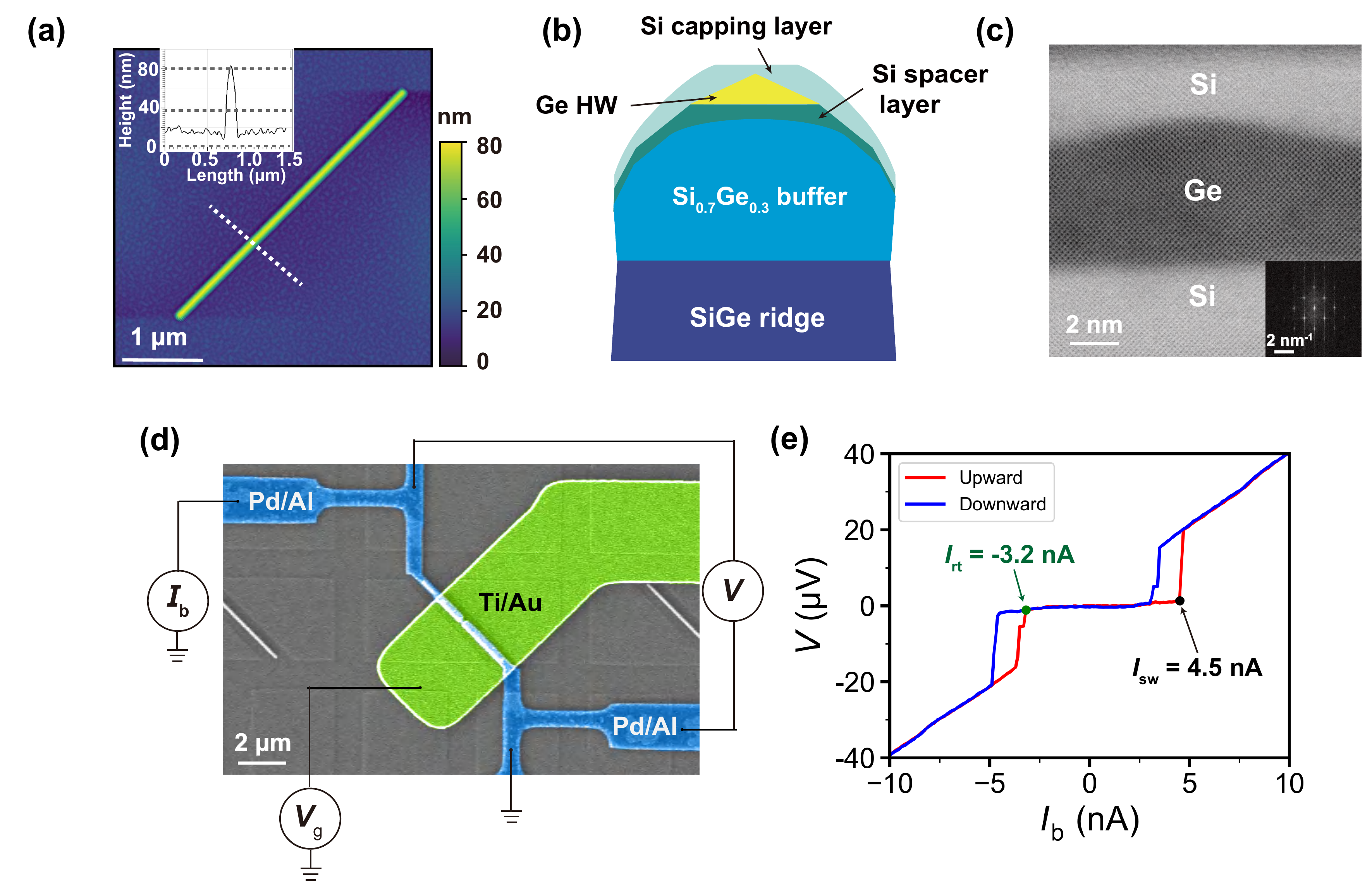}
\caption{Basic characteristics of the Josephson junction device made from a Ge hut nanowire grown by MBE. \textbf{(a)} AFM image of a Ge hut nanowire on top of a SiGe ridge. The inset shows a linecut across the nanowire and the height of the entire structure is $\sim70\,\mathrm{nm}$ measured from the post growth substrate surface. \textbf{(b)} Cross-sectional schematic of a Ge hut nanowire. The layer structures are marked with different colors and the Ge hut nanowire has a triangular cross section. \textbf{(c)} TEM image of a Ge hut nanowire structure. Different contrasts correspond to the regions with different elements: Ge (black) and Si (white). The inset shows the corresponding FFT image. \textbf{(d)} False-color SEM image of device D1 and corresponding measurement circuit in current-bias measurements. Superconducting electrodes are made from Pd/Al (blue) and the gate is made from Ti/Au (green). The junction length of the device is about $100\,\mathrm{nm}$. \textbf{(e)} Measured voltage $V_\text{}$ across the junction as a function of current bias $I_\text{b}$ at $V_\text{g}$ = $-0.5\,\mathrm{V}$. Red (blue) curve is measured in upward (downward) scanning direction. Switching current $I_\text{sw}$ and retrapping current $I_\text{rt}$ of the upward scanning trace are marked.
}\label{figure:1}
\end{figure*}

The Ge hut nanowires used in this work are grown on SiGe ridges via a low-temperature epitaxial method by MBE (see Ref.~\cite{Wang2025_NL} for details). Figure \ref{figure:1}a shows an atomic force microscopy (AFM) image of an as-grown Ge hut nanowire structure, and Figure \ref{figure:1}b schematically illustrates the layers grown in the structure. A $60\,\mathrm{nm}$ buffer layer of Si$_{0.7}$Ge$_{0.3}$ is initially grown on a prepatterned SiGe ridge of $\sim100\,\mathrm{nm}$ in width and $\sim60\,\mathrm{nm}$ in height on a SiGe/Si substrate, followed by the growth of a 4-nm Si spacer layer for surface planarization. Then, the Ge nanowire is grown on top, which exhibits a triangular cross section with a base width of $\sim65\,\mathrm{nm}$ and a height of $\sim7\,\mathrm{nm}$. Finally, a 3-nm Si capping layer is grown to prevent oxidation of the Ge nanowire. As seen in the inset of Figure \ref{figure:1}a, the height of the entire structure measured from the substrate surface after the growth of the nanowire structure is $\sim70\,\mathrm{nm}$. Figure \ref{figure:1}c shows a scanning transmission electron microscopy (STEM) image of a similarly grown Ge nanowire around its apex. It is seen that the interfaces between the Ge layer and surrounding Si layers are sharp, and there are no observable defects or dislocations in the nanowire. This indicates that a high-quality single-crystalline Ge nanowire is obtained. The inset shows the fast Fourier transform (FFT) pattern of Figure \ref{figure:1}c, and the single set of diffraction spots further confirms the exceptionally high crystalline quality of the nanowire. More details about the material characterization can be found in Ref.~\cite{Wang2025_NL}. 

Figure \ref{figure:1}d shows a false-color scanning electron microscope (SEM) image of a measured device (device D1) with the results presented in the main article and a schematic for the measurement circuit setup. The contact regions are defined by electron-beam lithography, followed by native oxide removal using a buffered oxide etch solution (BOE, 7:1 volume ratio) for 10 seconds. The BOE etching process is designed to completely remove the 1-2$\,\mathrm{nm}$ SiO$_{2}$ layer, leaving the underlying silicon layer intact. Subsequently, the superconducting electrodes are made by depositing $5/90\,\mathrm{nm}$ Pd/Al with electron-beam evaporation and a lift-off process. The distance between the two superconducting electrodes for device D1 is about $100\,\mathrm{nm}$. Then, $20\,\mathrm{nm}$ thick Al$_{2}$O$_{3}$ is grown by atomic layer deposition (ALD) at $300\,\mathrm{^\circ{C}}$ for a duration of 40 minutes, which also serves as an annealing process to form low-resistance contacts. Transversal diffusion of Pd/Al into the junctions might also happen and the length of the possible diffusion is believed to be substantially shorter than $100\,\mathrm{nm}$. Finally, $10/150\,\mathrm{nm}$ Ti/Au is evaporated to work as the top gate. After the fabrication, the devices are measured in a $^{3}$He/$^{4}$He dilution refrigerator equipped with a magnet. All measurements are performed at a base temperature of $40\,\mathrm{mK}$, unless otherwise specified. In this work, two similar devices (D1 and D2) are studied. All data presented in the main text are from D1, and the results of D2 are provided in the Supplementary Materials. The circuit in Figure \ref{figure:1}d is a typical quasi-four-terminal setup for current-bias measurements, where the current bias $I_\text{b}$ is applied while the voltage drop $V_\text{}$ across the junction is measured. The gate voltage $V_\text{g}$ is used to tune the semiconductor junction between the two superconducting electrodes. Figure \ref{figure:1}e displays the measured voltage $V_\text{}$ across the junction as a function of current bias $I_\text{b}$ at $V_\text{g}$ = $-0.5\,\mathrm{V}$. The red and blue traces correspond to $V_\text{}$-$I_\text{b}$ measurements by scanning $I_\text{b}$ upward and downward, respectively. Hysteresis is observed between the upward and downward scanning traces, which is commonly seen in nanostructured semiconductor Josephson junctions.\cite{Doh2005_Science,Nilsson2012_NanoLett,Wang2022_SciAdv,Levajac2023_NanoLett,Yan2023_NanoLett,Yan2025_AFM,Su_PhysRevLett,Wu2025_NJOP} Such a hysteretic behavior can be explained by phase instability in the junction and/or heating effects.\cite{Tinkham2004_introduction,Tinkham2003_PRB,Courtois2008_PRL} By taking the upward scanning trace, the switching current $I_\text{sw}$ and the retrapping current $I_\text{rt}$ are extracted as $4.5\,\mathrm{nA}$ and $-3.2\,\mathrm{nA}$, respectively. 

\begin{figure}[!t] 
\centering
\includegraphics[width=0.7\linewidth]{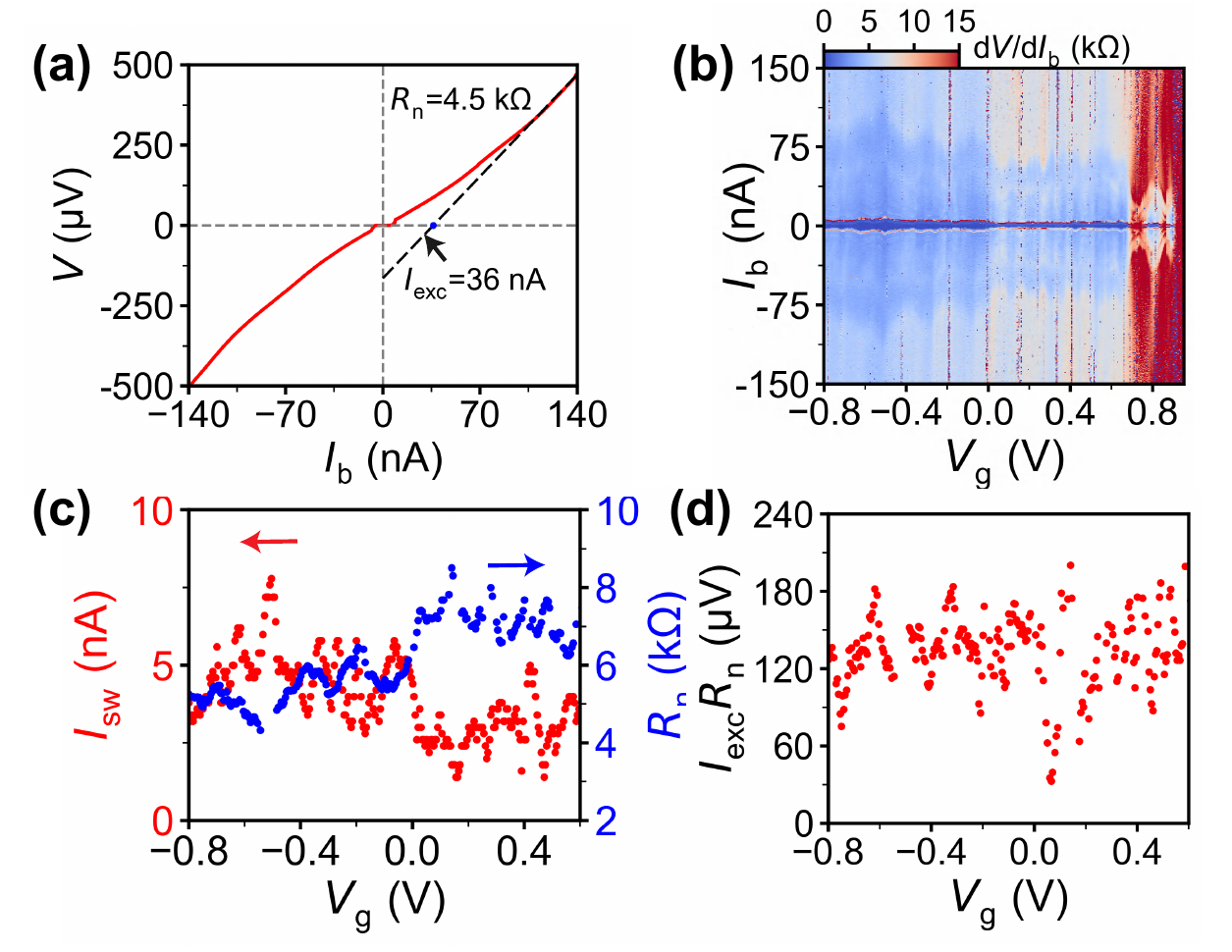}
\caption{Current-bias measurement results of device D1 at $T_\text{}$ = $40\,\mathrm{mK}$ and $B_\text{}$ = 0. \textbf{(a)} $V_\text{}$-$I_\text{b}$ curve at $V_\text{g}$ = $-0.5\,\mathrm{V}$. The black dashed line is the fitting curve of the linear region. The slope of this line corresponds to the normal state resistance $R_\text{n}$ and the intersection point between the line and the $I_\text{b}$ axis signifies the excess current $I_\text{exc}$. \textbf{(b)} Differential resistance d$V_\text{}$/d$I_\text{b}$ versus $I_\text{b}$ and $V_\text{g}$. \textbf{(c)} Switching current $I_\text{sw}$ (red dots) and normal state resistance $R_\text{n}$ (blue dots) versus $V_\text{g}$. The data are extracted from figure (b). \textbf{(d)} Product $I_\text{exc}$$R_\text{n}$ as a function of $V_\text{g}$. The data of $I_\text{exc}$ and $R_\text{n}$ are extracted from figure (b). The value of the product ranges from 50 to $200\,\mathrm{\mu V}$.
}\label{figure:2}
\end{figure}

Figure \ref{figure:2}a shows the $V_\text{}$-$I_\text{b}$ curve in a large range of $I_\text{b}$ at $V_\text{g}$ = $-0.5\,\mathrm{V}$. As shown in the figure, dissipationless supercurrent with zero resistance can be observed at low $I_\text{b}$. In this case, the entire Josephson junction stays in the superconducting state induced by proximity effect. At larger $I_\text{b}$ ($I_\text{b}$ > $100\,\mathrm{nA}$), $V_\text{}$ across the junction exceeds 2$\Delta$/$e$ with $\Delta$ being the superconducting gap, and $V_\text{}$ exhibits a linear dependence on $I_\text{b}$. By extrapolating the linear range of the trace, excess current $I_\text{exc}$ arising from dissipationless Cooper-pair transport\cite{Octavio1983_PRB,Flensberg1988_PRB} and normal-state resistance $R_\text{n}$ can be obtained. The extracted values of $I_\text{exc}$ and $R_\text{n}$ are $36\,\mathrm{nA}$ and $4.5\,\mathrm{k\Omega}$, respectively. Figure \ref{figure:2}b presents the color map of differential resistance d$V_\text{}$/d$I_\text{b}$ as a function of $I_\text{b}$ and $V_\text{g}$. The dark blue region at the center of the diagram represents the superconducting state and the faint fringes at higher $I_\text{b}$ are caused by MAR process. It is obvious that the performance of the Josephson junction is significantly dependent on $V_\text{g}$. At $V_\text{g}$ > $0.7\,\mathrm{V}$, the junction is in a relatively close regime and the supercurrent is hardly detectable in this regime. At $V_\text{g}$ < $0.7\,\mathrm{V}$, the junction is in a relatively open regime. In this regime, there is gate-tunable supercurrent and the differential resistance at voltage bias above 2$\Delta$/$e$ fluctuates with the gate voltage. The fluctuation can be attributed to the formation of randomly distributed quantum dots\cite{Levajac2024_PRL} or the emergence of Fabry-Pérot interference in the semiconductor junction.\cite{Liang2001_Nature,Kretinin2010_NL}

\begin{figure}[!t] 
\centering
\includegraphics[width=0.7\linewidth]{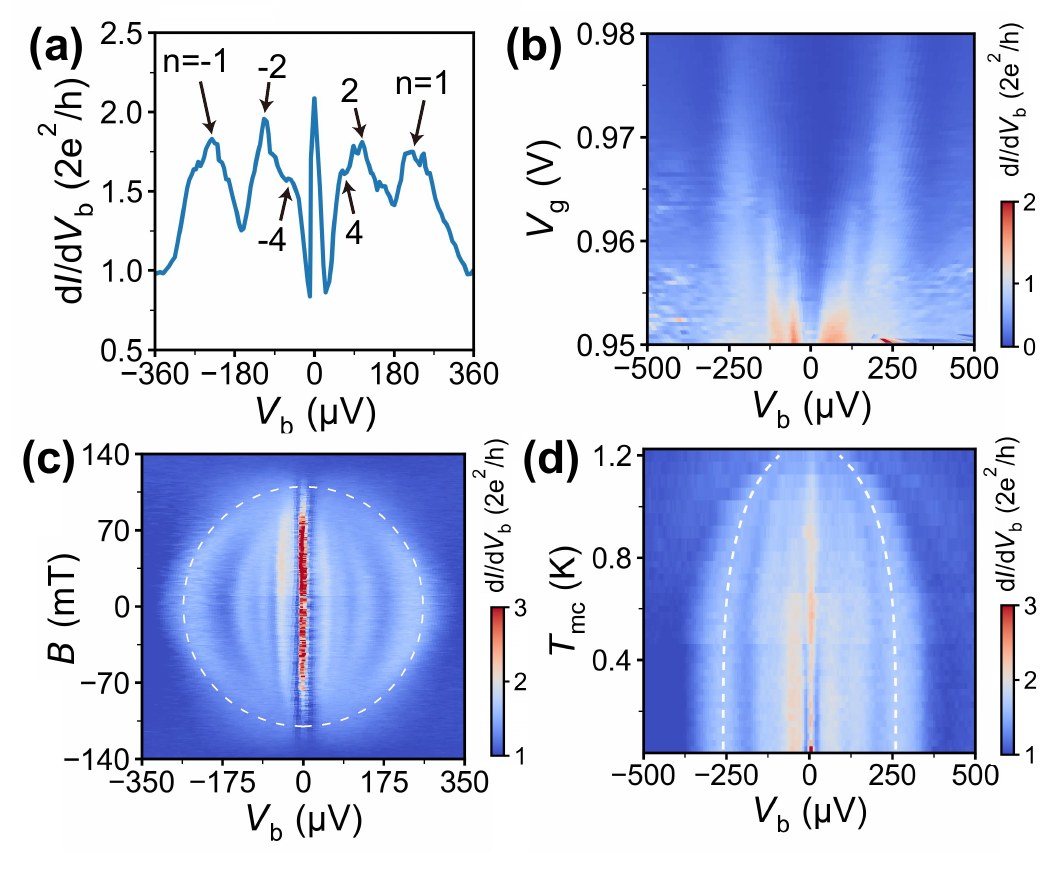}
\caption{Voltage-bias measurements of device D1. \textbf{(a)} Differential conductance d$I_\text{}$/d$V_\text{b}$ as a function of voltage bias $V_\text{b}$ at $V_\text{g}$ = $0.828\,\mathrm{V}$ and $B_\text{}$ = 0. The conductance peaks marked at finite $V_\text{b}$ are due to MAR processes. The numbers n=$\pm{1}$, $\pm{2}$, $\pm{4}$ indicate the orders of the MARs. \textbf{(b)} Color map of d$I_\text{}$/d$V_\text{b}$ versus $V_\text{b}$ and $V_\text{g}$. High-order MAR peaks show up at lower $V_\text{g}$ as a result of more opened channels for hole transport. \textbf{(c)} Color map of d$I_\text{}$/d$V_\text{b}$ as a function of $V_\text{b}$ and magnetic field $B_\text{}$ at $V_\text{g}$ = $0.822\,\mathrm{V}$ and base temperature. The magnetic field is applied in the plane of the substrate, perpendicular to the axis of the Ge hut nanowire. The critical field $B_\text{c}$ is $\sim0.12\,\mathrm{T}$. \textbf{(d)} Temperature-dependent results of voltage-bias measurements at $V_\text{g}$ = $0.812\,\mathrm{V}$ and $B_\text{}$=0. The critical temperature $T_\text{c}$ is $\sim1.4\,\mathrm{K}$. The white dashed lines in (c) and (d) are fitting curves of the first-order MAR peaks according to the BCS theory.
}\label{figure:3}
\end{figure}

Figure \ref{figure:2}c presents the gate-voltage dependence of $I_\text{sw}$ and $R_\text{n}$, extracted from Figure \ref{figure:2}b. Here, the gate voltage is restricted to $V_\text{g}$ < $0.6\,\mathrm{V}$, maintaining a finite supercurrent flow. In such a gate voltage range, $I_\text{sw}$ varies from 1.2 to $7.8\,\mathrm{nA}$ while $R_\text{n}$ fluctuates in the range of 4.3 to $8.5\,\mathrm{k\Omega}$. The value of the product $I_\text{sw}$$R_\text{n}$ varies in the range of 11 to $37\,\mathrm{\mu V}$. The value of the product extracted for device D2 is found to vary with gate and can be even lower than $21.3\,\mathrm{\mu V}$ (see Figure S1 in the Supplementary Materials). According to theory, the product $I_\text{c}$$R_\text{n}$ of short dirty Josephson junctions, where $I_\text{c}$ is the critical current, should have a value of $\sim{\pi\Delta/}{2e}$.\cite{Beenakker1991_PRL} For comparison, the product $I_\text{sw}$$R_\text{n}$ of our device has a lower value than the predicted value $\pi\Delta/{2e}$ $\sim200\,\mathrm{\mu V}$, by taking $\Delta=130\,\mathrm{\mu eV}$ (see Figure \ref{figure:3}). This can be attributed to premature switching of the Josephson junctions,\cite{Tinkham2004_introduction} causing a lower value of $I_\text{sw}$ compared with $I_\text{c}$. Figure \ref{figure:2}d shows the product $I_\text{exc}$$R_\text{n}$ as a function $V_\text{g}$. The value of the product varies from 50 to 200$\,\mathrm{\mu V}$. The fluctuations of the value are likely due to the presence of resonant states and/or unintentional dots, and the peaks correspond to the case that the junction is on resonance and Cooper-pair transport is therefore enhanced. According to Blonder-Tinkham-Klapwijk (BTK) model and following theory,\cite{Blonder1982_PRB,Octavio1983_PRB,Flensberg1988_PRB} the transparency of a Josephson junction follows the form $T_\text{r}$ = 1/(1+Z$^{2})$ and the parameter Z is related to the ratio $e$$I_\text{exc}$$R_\text{n}$/$\Delta$. In our device, the ratio $e$$I_\text{exc}$$R_\text{n}$/$\Delta$ is in the range of 0.38 to 1.54, leading to junction transparency from 57\% to 85\%. The junction transparency for device D2 is slightly lower, ranging from 46\% to 76\% (see Figure S1 in the Supplementary Materials). The value of the ratio $e$$I_\text{exc}$$R_\text{n}$/$\Delta$ indicates that our Ge hut nanowire Josephson junction devices are likely in a short dirty regime,\cite{Artemenko1979_ZETF} which is consistent with the analysis of Figure \ref{figure:2}c. This might be due to an inhomogeneous Ge-Al interface in our devices as a result of an uncontrollable annealing process during device fabrication.

Figure \ref{figure:3} shows the voltage-bias measurements of device D1. Figure \ref{figure:3}a presents the differential conductance d$I_\text{}$/d$V_\text{b}$ in the unit of 2$e$$^{2}$/h as a function of voltage bias $V_\text{b}$ at $V_\text{g}$ = $0.828\,\mathrm{ V}$ and $B_\text{}$ = 0. The conductance peaks at finite $V_\text{b}$ are characteristic fingerprints of MARs, where quasiparticles undergo coherent scattering at the two superconductor-semiconductor interfaces.\cite{Octavio1983_PRB} Conductance peaks originating from MARs appear at $V_\text{b}$ = 2$\Delta$/$(ne)$, where n denotes the order of the MAR peaks and $\Delta$ is the superconducting gap. As seen in the figure, MARs up to the fourth order are observed. The superconducting gap is extracted to be $\sim130\,\mathrm{\mu eV}$. Figure \ref{figure:3}b shows the differential conductance d$I_\text{}$/d$V_\text{b}$ as a function of $V_\text{b}$ and $V_\text{g}$. At high $V_\text{g}$, only the first-order MAR peaks are observable. With decreasing $V_\text{g}$, high-order MAR peaks gradually show up as a result of the opened hole channel in the junction. After that, the evolution of the MAR peaks with magnetic field and temperature is studied. Figure \ref{figure:3}c displays d$I_\text{}$/d$V_\text{b}$ as a function of $V_\text{b}$ and magnetic field $B_\text{}$ at $V_\text{g}$ = $0.822\,\mathrm{V}$. The magnetic field is applied in the plane of the substrate, perpendicular to the axis of the Ge hut nanowire. Increasing magnetic field leads to a monotonic shift of the MAR peaks towards lower $V_\text{b}$, culminating in complete suppression beyond a certain field. According to the Bardeen-Cooper-Schrieffer (BCS) theory, the superconducting gap in a magnetic field follows the form of $\Delta(B)=\Delta(0)[1-(B/B_c)^2]^{1/2}$.\cite{Tinkham2004_introduction} The fitting trace (white dashed line) overlaps well with the first-order MAR peaks by taking $\Delta(0)=130\,\mathrm{\mu eV}$. The critical field $B_\text{c}$ is extracted as $\sim0.12\,\mathrm{T}$. Similar measurements have been performed on device D2, and comparable results have been obtained (see Figure S2 in the Supplementary Materials). Figure \ref{figure:3}d shows the temperature dependence of the MAR peaks at $V_\text{g}$ = $0.812\,\mathrm{V}$. As expected, the MAR peaks are squeezed to zero bias voltage with increasing temperature as a result of suppressed superconductivity by temperature. The temperature dependence of superconducting gap can be approximately described by an empirical formula $\Delta(T)=\Delta(0)\sqrt{cos[(\pi/2)(T/T_{c})^2]}$,\cite{Thomas1966_PR,Nilsson2012_NanoLett} where $T_{c}$ is the critical temperature. The white dashed lines indicate the predicted dependence of $2\Delta(T)/e$ on temperature from the formula. The critical temperature $T_{c}$ is extracted to be $\sim1.4\,\mathrm{K}$. 

In summary, we have presented the first experimental study of Josephson junctions fabricated from Ge hut nanowires. Both the supercurrent and the excess current exhibit substantial dependence on the gate voltage. The transparency of the Josephson junctions fluctuates with gate voltage and can be as high as 85\%. In voltage-bias spectroscopy measurements, distinct MARs up to the fourth order are resolved. With magnetic field- and temperature-dependent measurements of the MARs, the critical magnetic field and the critical temperature of the induced superconductivity in our fabricated Ge hut nanowire Josephson junctions are extracted to be $\sim0.12\,\mathrm{T}$ and $\sim1.4\,\mathrm{K}$, respectively. Our results demonstrate that decent induced superconductivity has been achieved in Ge hut nanowires. Combined with the intrinsic advantages of Ge hut nanowires, this work highlights the great potential of the system for exploring exotic quantum phenomena and building advanced quantum devices.

\section{Supplementary Material}

In the Supplementary Material, we have provided data of device D2, including (1) supercurrent, excess current and normal resistance as a function of gate voltage; (2) multiple Andreev reflections as a function of gate and magnetic field.

\section{Author contributions}

H.Q.X conceived and supervised the project. H.G. and J.H.W fabricated the devices. J.H.W. and J.J.Z grew the Ge hut nanowires. H.G. and J.Y.W. performed the electrical transport measurements. H.G., J.Y.W. and H.Q.X. analyzed the measurement data. H.G., J.Y.W. and H.Q.X. wrote the manuscript with inputs from all the authors.  
\begin{acknowledgement}
This work was supported by the National Natural Science Foundation of China (Grant Nos. 92165208, 12374480, 12304101, 62225407 and 92165207) and the Innovation Program for Quantum Science and Technology (No. 2021ZD0302300). 
\end{acknowledgement}

\section{Data availability} 
The data that support the findings of this study are available from the corresponding authors upon reasonable request.  
      
\section{Conflict of interests}
The authors declare no conflict of interests. 

\bibliography{references}

\newpage

\end{document}


\begin{CJK}{UTF8}{gbsn} 

\clearpage
\section{Additional Data} 

\begin{figure*}[!h]
\centering
\includegraphics[width=1\linewidth]{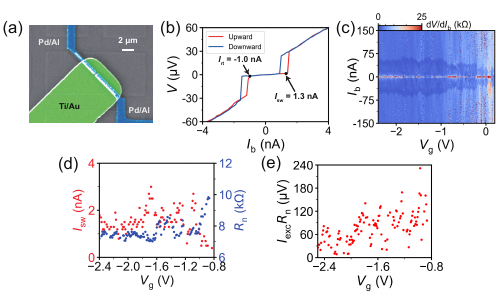} 
\caption{\doublespacing \textbf{False-color SEM image and current-bias measurements of device D2.} \textbf{(a)} False-color SEM image of device D2. Superconducting electrodes are made from Pd/Al (blue) and top gate is made from Ti/Au (green). The junction length of the device is $\sim120\,\mathrm{nm}$. \textbf{(b)} Measured voltage $V_\text{}$ across the junction as a function of current bias $I_\text{b}$ for both upward and downward current sweep directions at $V_\text{g}$ = $-1.91\,\mathrm{V}$. For the upward scanning trace, the switching current $I_\text{sw}$ and the retrapping current $I_\text{rt}$ are extracted as $1.3\,\mathrm{nA}$ and $-1.0\,\mathrm{nA}$, respectively. \textbf{(c)} Differential resistance d$V_\text{}$/d$I_\text{b}$ versus $I_\text{b}$ and $V_\text{g}$. \textbf{(d)} Switching current $I_\text{sw}$ (red dots) and normal state resistance $R_\text{n}$ (blue dots) as a function of $V_\text{g}$. The data are extracted from figure (c). The value of product $I_\text{sw}$$R_\text{n}$ varies from 3.1 to $21.3\,\mathrm{\mu V}$. \textbf{(e)} Product $I_\text{exc}$$R_\text{n}$ as a function of $V_\text{g}$. The data of $I_\text{exc}$ and $R_\text{n}$ are extracted from figure (c). The value of the product ranges from \textcolor{red}{8.67} to \textcolor{red}{$169.4\,\mathrm{\mu V}$} leading to the transparency of this Josephson junction to be \textcolor{red}{46\%} to \textcolor{red}{76\%} by taking $\Delta$ = $170\,\mathrm{\mu eV}$. The superconducting gap of the device is adopted from Fig. S2.
}\label{figure:S1}
\end{figure*}

\clearpage

\begin{figure*}[!h]
\centering
\includegraphics[width=1\linewidth]{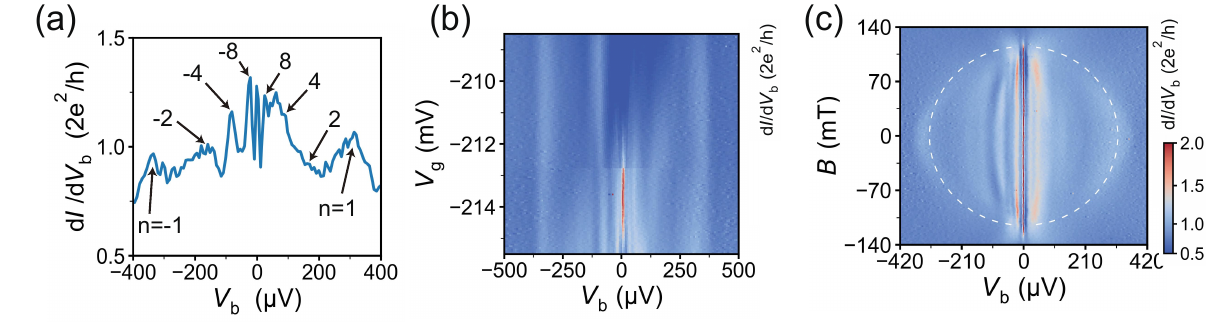} 
\caption{\doublespacing \textbf{Voltage-bias measurement results of device D2.} \textbf{(a)} Differential conductance d$I_\text{}$/d$V_\text{b}$ as a function of $V_\text{b}$ at $V_\text{g}$ = $-215.2\,\mathrm{mV} $ and $B_\text{}$ = 0. The numbers n=$\pm{1}$, $\pm{2}$, $\pm{4}$, $\pm{8}$ mark the orders of the MARs. \textbf{(b)} The color map of d$I_\text{}$/d$V_\text{b}$ versus $V_\text{b}$ and $V_\text{g}$. \textbf{(c)} Differential conductance d$I_\text{}$/d$V_\text{b}$ as a function of $V_\text{b}$ and magnetic field $B_\text{}$. The white dashed line is the fitting curve of the magnetic field dependence of the first-order MAR peaks according to the BCS theory. The extracted superconducting gap $\Delta$ and critical field $B_\text{c}$ are $\sim170\,\mathrm{\mu eV}$ and $\sim125\,\mathrm{mT}$, respectively.
}\label{figure:S2}
\end{figure*}

\clearpage

\end{CJK}